\documentclass[conference]{IEEEtran}

\IEEEoverridecommandlockouts
\usepackage{cite}
\usepackage{amsmath,amssymb,amsfonts}
\usepackage{graphicx}
\usepackage{textcomp}
\usepackage{xcolor}
\usepackage{epsfig}
\usepackage{color}
\usepackage{multirow}
\usepackage{CJK}
\usepackage{algorithm,algpseudocode}
\usepackage{subfig}

\usepackage[left=1.50cm, right=1.50cm, top=1.85cm, bottom=2.1cm]{geometry}
\usepackage{amsmath} 
\floatname{algorithm}{Algorithm}

\def\BibTeX{{\rm B\kern-.05em{\sc i\kern-.025em b}\kern-.08em
    T\kern-.1667em\lower.7ex\hbox{E}\kern-.125emX}}
\begin{document}

\title{Graph Sparsification with Generative Adversarial Network}
\author{\IEEEauthorblockN{Hang-Yang Wu}
\IEEEauthorblockA{\textit{Department of Computer Science and Information Eng.} \\
\textit{National Taiwan University of Science and Technology}\\
Taipei, Taiwan \\
M10615096@mail.ntust.edu.tw}
\and
\IEEEauthorblockN{Yi-Ling Chen}
\IEEEauthorblockA{\textit{Department of Computer Science and Information Eng.} \\
\textit{National Taiwan University of Science and Technology}\\
Taipei, Taiwan \\
yiling@mail.ntust.edu.tw}
}
\maketitle

\begin{abstract}
Graph sparsification aims to reduce the number of edges of a network while maintaining its accuracy for given tasks. In this study, we propose a novel method called GSGAN, which is able to sparsify networks for community detection tasks. GSGAN is able to capture those relationships that are not shown in the original graph but are relatively important, and creating artificial edges to reflect these relationships and further increase the effectiveness of the community detection task. We adopt GAN as the learning model and guide the generator to produce random walks that are able to capture the structure of a network. Specifically, during the training phase, in addition to judging the authenticity of the random walk, discriminator also considers the relationship between nodes at the same time. We design a reward function to guide the generator creating random walks that contain useful hidden relation information. These random walks are then combined to form a new social network that is efficient and effective for community detection. 
Experiments with real-world networks demonstrate that the proposed GSGAN is much more effective than the baselines, and GSGAN can be applied and helpful to various clustering algorithms of community detection.
\end{abstract}


\section{Introduction}
There are numerous types of data that can be represented by graphs. For example, social networks, financial transaction networks, communication networks, and citation networks. As the size of data grows rapidly, it becomes impractical to analyze, store, and visualize such a large amount of network data. Thus, the technique of sparsification becomes more and more important.
Graph sparsification can bring the following benefits: 1) After being sparsified, the graph size becomes smaller, and the storage space is thereby reduced. 2) Certain graph algorithms are not able to handle large graphs due to the high time complexity. However, if the graph is sparsified first, we can save considerable computation time and maintain the accuracy when applying these algorithms. 3) Graph sparsification can eliminate redundant data and noises from the graph. In addition to improving the performance of graph analysis algorithms, sparsification also makes the graph easier to be visualized. 4) When there are privacy concerns with the data, sparsification can remove and hide certain information from the graph to achieve better privacy protection.

The subsequent applications of graph sparsification can be diverse, leading to different requirements of sparsification, such as retaining specific structural patterns, paying attention to certain network entities, preserving graph query answers or maintaining the distribution of graph attributes. In general, the design of graph sparsification algorithms may involve the following four issues: 
\begin{figure}[htbp]
    \centering
    \includegraphics[width=0.4\textwidth]{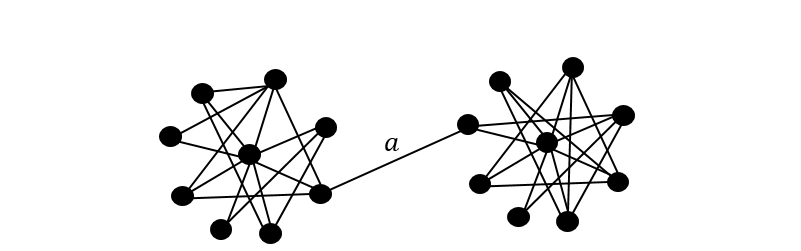}
    \caption{A simple example of different tasks requiring different edges.}
    \label{fig:fig1}
\end{figure}
1) Different subsequent tasks favor different kinds of sparsified graphs. Fig. \ref{fig:fig1} is an illustrative example. For the influence maximization task, edge $a$ should be preserved in the sparsified graph; however, $a$ could be a redundant edge for the community detection task. Therefore, the design of sparsification algorithms  needs to take the subsequent tasks into consideration. 
2) One of the main purposes of graph sparsification is to reduce the size of large input graphs. Therefore, the sparsification algorithms need to be capable of handling input graphs with large sizes. 
3) In the real world, networks usually contain lots of noises that affect the performance of graph analysis. Therefore, it is also important for the sparsification algorithm to eliminate these noises, such that the performance of subsequent tasks can be further improved. 
4) Previous studies of graph sparsification \cite{satuluri2011local}\cite{hamann2016structure}\cite{gionis2017community} usually focus on existing edges in the graph and make selections among them. However, this can lead to a key question: \emph{\textbf{What if the edges not shown in the original graph contain more important information?}} For certain network analysis tasks, creating artificial edges for better results is allowed. If the perspective of generation can be incorporated (instead of using only the perspective of elimination like the traditional methods), we will be able to generate new edges that do not exist in the original network but are essential for the subsequent tasks.

\begin{figure}[htpb]
    \centering
    \includegraphics[width=0.33\textwidth]{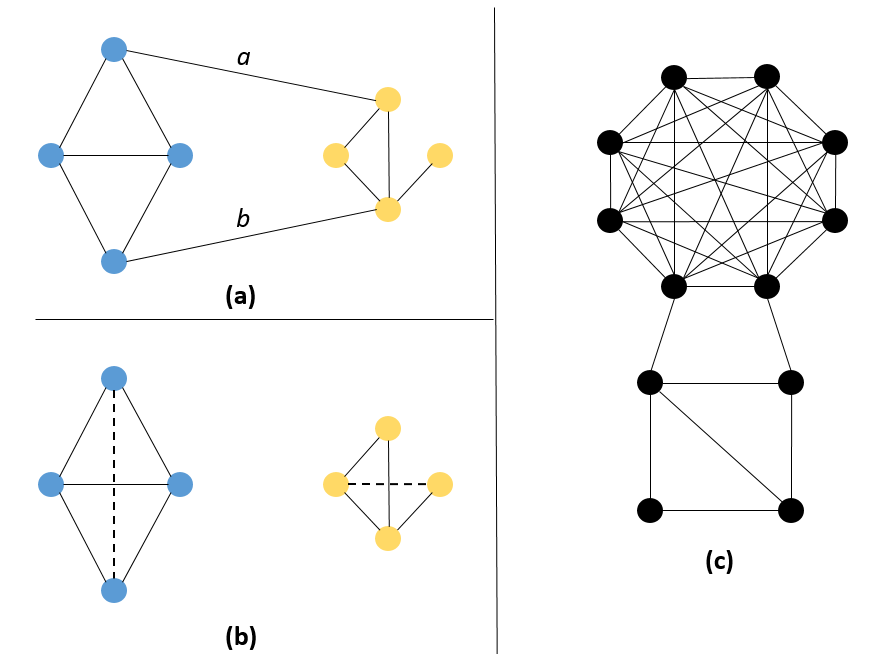}
    \caption{(a) and (b) provide a comparison of two different graphs for the community detection task. (c) is an example illustrating different communities with different densities.}
    \label{fig:fig2}
\end{figure}

\textbf{Example 1:}
Here we provide an example to better illustrate issue 4. Fig. \ref{fig:fig2}(a) shows an example graph, where the blue nodes and yellow nodes represent different communities. If we replace some existing edges between different communities (e.g., edges $a$ and $b$) with some artificial edges inside the community (e.g., the dotted lines in Fig. \ref{fig:fig2}(b)), the community structure would become more compact under the same number of edges, which is conducive to the community detection task. In the traditional sparsification methods, this kind of artificial edges are not considered; by contrast, the proposed method in this study is able to create artificial edges that do not exist in the original network but contain important information for the given task during the graph sparsification.

In this study, we propose a novel graph sparsification method called Graph Sparsification with Generative Adversarial Network (GSGAN) to address the above four issues. 
For the issue 1, GSGAN can flexibly replace components (i.e., reward functions) to provide different directions of sparsification that fit the subsequent tasks. 
For the issue 2, GSGAN divides the network into multiple random walks, in order to be capable of processing huge data. 
For the issue 3, GSGAN can accurately filter out noises and focus on retaining important information in the sparsified graph. 
For the issue 4, GSGAN is able to capture those relationships that are not shown in the original network but are relatively important, and creating artificial edges to further increase the effectiveness of the community detection task. 
We use GAN \cite{goodfellow2014generative} as the learning model and guide Generator $G$ to produce random walks that capture the structure of the network. Specifically, during the training process, in addition to judging the authenticity of the random walk, Discriminator $D$ also considers the relationship between nodes at the same time. We design a reward function to guide the generator creating random walks that contain useful hidden relation information. These random walks are then combined to form a new social network that is efficient and effective for community detection. 
Experiments with real-world networks demonstrate that the proposed GSGAN is much more effective than the baselines, and GSGAN can be applied and helpful to various clustering algorithms of community detection.

The main contributions of this study can be summarized as follows:
\begin{itemize}
\item We observe that previous studies only focus on the existing edges and make selections from them, but for certain scenarios, such as community detection, it is allowed to create additional edges for better results. Therefore, our GSGAN is designed to be able to create artificial edges and further increase the effectiveness of community detection tasks.

\item 
We compare the proposed GSGAN with other baselines with various network datasets and different clustering algorithms of community detection. 
The experimental results show that the sparsified graphs generated by GSGAN lead to community detection results significantly superior to the sparsified graphs obtained from the baseline sparsification algorithms. 
It is worth noting that GSGAN is able to handle a much wider range of ratios than other baseline methods. When the ratio is low (e.g., 1\%), only the sparsified network generated by GSGAN is still capable of reaching acceptable community detection results. 

\item 
Despite of only using far fewer edges (e.g., 5\%), applying clustering algorithms of community detection on our sparsified graph receives comparable or even better results than on the original graph. 
Moreover, the execution time of community detection can be considerably reduced (e.g., by nearly an order of magnitude) when applying on the sparsified graph. 

\end{itemize}
The rest of the paper is organized as follows. Section 2 surveys the related works. Section 3 defines the problem and introduces the proposed GSGAN. Section 4 shows the experimental setup and results. Finally, we conclude our study in Section 5.

\section{Related Works}
In this section, we introduce and discuss some previous works related to graph sparsification. Moreover, we also introduce the studies related to network generation, since our sparsification algorithm also generates some important artificial edges. 

\subsection{Graph Sparsification}
Sparsification has been extensively studied. This technology is applied to different domains, such as community detection, influence maximization and retaining the properties of the network. According to different goals, this problem can be roughly divided into two categories: 1) Preserving network properties. 2) Application-based sparsifiers.

\subsubsection{Preserving network properties}
Many studies focus on preserving properties of the network, such as shortest path distances, graph cuts, laplacian, diameter. In \cite{peleg1989graph}, the authors aim at preserving the shortest path distance between all node pairs is based on the concept of t-spanner. Cut-based sparsification preserves the size of all cuts of the graph within a multiplicative error. In \cite{fung2019general}, the authors simplify analysis by utilizing sampling probabilities inversely proportional to the size of the minimum cut separating $u$ and $v$. In \cite{nagamochi1992linear}, the authors estimate the edge connectivities using the NI index $\lambda_e$ of an edge $e$. In \cite{spielman2011graph}, the authors generate the electrical equivalent of the graph by assuming resistors of 1omega at each link. The sampling probability of $edge (u, v)$ is proportional to the amount of current that flows through e when a unit voltage difference is applied to $u$, $v$. In \cite{john2016single}, the authors use the algebraic distance between two nodes to measure the sparseness, and maintain the network properties as much as possible, PageRank, modularity, diameter, degree centrality, etc.

\subsubsection{Application-based sparsifiers}
Another type of graph sparsification is to do sparsification for specific tasks, such as community detection, influence maximization, link prediction. It is worth noting that different tasks require different graph sparsification. For link prediction, DEDS \cite{chen2015ensemble} processes the original graph into multiple smaller networks to improve the efficiency of link prediction. SPINE \cite{mathioudakis2011sparsification} mainly do sparsification for influence maximization. It uses a maximum likelihood method to find a set of edges that generates an observed set of influence cascades with the greatest probability. L-Spar \cite{satuluri2011local} develops a sparsification algorithm specifically for graph clustering. The underlying principle is to preferentially retain edges between nodes that are in the same cluster by selecting edges between nodes with similar neighbors. LD \cite{hamann2016structure} proposes some heuristic methods aiming to preserve the structure of social networks. Their main method Local Degree keeps only the edges to hubs, i.e., vertices with high degree, claiming that they are crucial for the topology of complex social networks. NETSPARSE \cite{gionis2017community} proposes a new formulation of network sparsifcation, where the input consists not only of a network but also of a set of communities.

Our proposed GSGAN seems more application-based, because it is currently designed for the task of community detection. However, GSGAN has the flexibility of replacing components (i.e., reward functions) to handle different sparsified directions. 

\subsection{Network Generation}
Typically, a social network is represented by the adjacency matrix. In \cite{tavakoli2017learning}, the authors apply GAN \cite{radford2015unsupervised} to graph data by trying to directly generate an entire matrix. However, it is not suitable to use deep learning to directly generate an entire matrix, which is limited by the size of the matrix. To address this limitation, NETGAN \cite{bojchevski2018netgan} tends to capture the topology of the social network by random walks. Intuitively, NETGAN already has the effect of preserving important edges, since the edges of the random walks are sampled according to their frequency when the random walks are combined to form a whole social network. Theoretically, more important edges will appear more often in random walks. Therefore, using NETGAN architecture directly may also produce sparsified graph $G'$, but there are no guarantees that this $G'$ will be effective for the analysis of community detection. Inspired by NETGAN, we use random walks to do graph sparsification. 

\section{Approach}
In Section 3.1, we first define the problem. Section 3.2 then introduces our GSGAN model, and Section 3.3 explains the training process. In Section 3.4, we introduce the proposed measurements, and Section 3.5 details how to use GSGAN model to form the sparsified network.

\begin{figure*}[htbp]
    \centering
    \includegraphics[width=1\textwidth]{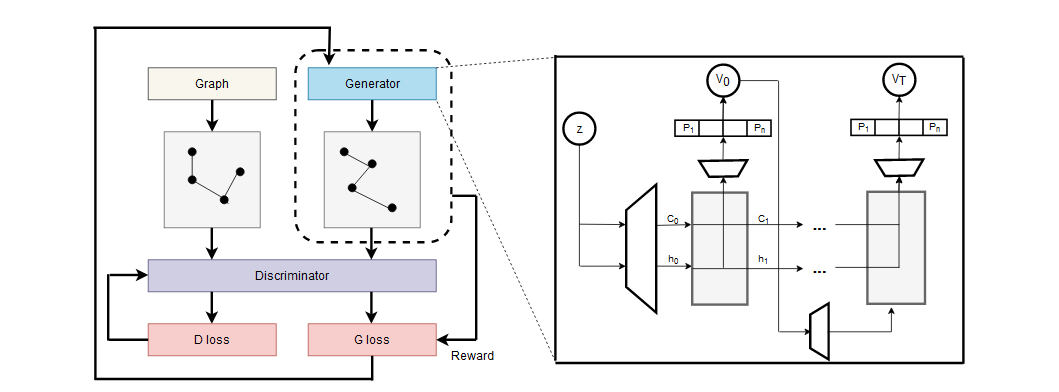}
    \caption{The architecture of GSGAN and the inside structure of the generator.}
    \label{fig:fig7}
\end{figure*} 

\subsection{Problem Definition}
Given a static, undirected graph $G(V, E)$, where $V$ represents the set of nodes in the graph and $E$ represents the set of edges in the graph. The objective of GSGAN is to find a sparsified graph $G'(V, E')$ with fewer edges, while the quality of community detection results is retained or even improved.

Please note that in the sparsified graph $G'$, we may add some new and important artificial edges. One important contribution of this study is that we address the issue illustrated in Example 1 by introducing the artificial edges. As show in the experiments, these artificial edges are quite helpful for community detection. To the best of our knowledge, this artificial edge mechanism has not been considered in previous studies of graph sparsification. 

\subsection{Model}
In this subsection, we provide the details of the proposed method, Graph Sparsification with Generative Adversarial Network (GSGAN). GSGAN consists of three components: 1) GAN, 2) LSTM and 3) Policy Gradient. We use GAN \cite{radford2015unsupervised} as the learning model to learn random walks that conform to the network structure of the original network. As the random walk contains an ordered sequence of nodes, both the generator and the discriminator in GAN use LSTM \cite{sundermeyer2012lstm} structure with time series memory. 
In addition, we also apply policy gradient to guide our model towards a specific direction through reward, then performing graph sparsification for community detection. In the following, we introduce the proposed GSGAN and its components in a bottom-up way.

\underline{\textbf{GAN:}}
The general idea of GAN is to simultaneously train a generator $G$ and a discriminator $D$ with conflict objectives. The generator $G$ tries to capture data distribution of the training set, and the discriminator $D$ tries to identify samples produced by the generator from real ones. That is, $G$ and $D$ play the following two-player min-max game:

\begin{equation}
\begin{aligned}
\min _{G} \max _{D} V(D, G) &=\mathbb{E}_{\boldsymbol{x} \sim p_{\text {data }}(\boldsymbol{x})}[\log D(\boldsymbol{x})] \\
&+\mathbb{E}_{\boldsymbol{z} \sim p_{\boldsymbol{z}}(\boldsymbol{z})}[\log (1-D(G(\boldsymbol{z})))]
\end{aligned}
\end{equation}

\underline{\textbf{LSTM:}}
LSTM is controlled by input and output gates which determine whether to update the hidden layer state. While RNN \cite{mikolov2010recurrent} only has a single transmission state, LSTM has two cell states $c^t$ and $h^t$, where $c^t$ changes slowly and $h^t$ changes drastically. The output $y^t$ is determined by $c^{t-1}$, $h^{t-1}$ and the current input.

\begin{equation}
c^{t} = z^{f} \odot c^{t-1} + z^{i} \odot z
\end{equation}
\begin{equation}
h^{t} = z^{o} \odot tanh(c^{t})
\end{equation}
\begin{equation}
y^{t} = \sigma (Wh^{t})
\end{equation}

\underline{\textbf{Policy gradient:}}
Reinforcement learning guides the learning goals through the
rewards obtained from the interaction between action and
environment. Policy gradient (PG) \cite{dayan2002reward} is a method to achieve reinforcement learning. The core concept of PG is to maximize the rewards obtained by performing action $a_{i} = \{ a_{1}, a_{2}, ..., a_{n} \}$ in the environment $s_{i} = \{ s_{1}, s_{2}, ..., s_{n} \}$.
Before updating the network parameters, the reward of the sample will be calculated in advance. When updating the network parameters, the reward is considered as the weight of the loss. 

\begin{equation}
\begin{aligned}
\nabla \bar{R}_{\theta} &= \sum_{\tau} R(\tau) \nabla p_{\theta}(\tau)\\
&= \mathbb{E}_{(a_{t}|s_{t}) \sim p_{\theta}(a_{t}|s_{t})}[R(a_{t}|s_{t}) \nabla \log p_{\theta}(a_{t}|s_{t})]
\end{aligned}
\end{equation}

\underline{\textbf{GSGAN:}}
An overview of GSGAN architecture and the internal structure of generator are provided in Fig. \ref{fig:fig7}. Here we use WGAN \cite{arjovsky2017wasserstein} instead of the traditional GAN architecture, because WGAN is more effective during the learning process. The interior of the generator is mainly the LSTM structure. The number of LSTM layers is the same as the length of the random walks to be generated. For the generation process, a latent vector $z$ drawn from a normal distribution is passed through a dense matrix $W \in \mathbb{R}$ to get initial values $(c^{0}, h^{0})$.

At each time step $t$, to generate the next node in the random walk, the output of LSTM cell will refer to $c^{t-1}$, $h^{t-1}$ and the previous node, and then map to a probability vector $p_{t} = \{p_{1}, p_{2}, ..., p_{n}\}$ through a fully connected dense $W$, where $W \in \mathbb{R}^{d \times N}$. We select the node corresponding to the maximum value in $p_{t}$ as the current node $v_{t}$. The above process repeats until the random walk reaches the required length. After the random walk is generated, we update network parameters with the policy gradient. The social network graph is considered as the environment, and the generation of edges in the random walk is considered as actions. We score the importance for each edge in the random walk and sum up these scores as a reward for generating this random walk. Let $RW$, $T$, $SN$ and $e$ denote the random walk, the length of random walk, the social network and the edge of random walk, respectively, then the reward can be formulated as:

\begin{equation}
f_{reward}(RW) = 
  \begin{cases}
    \sum_{i=1}^{T} f_{score}(SN, e_{i})    & \quad \text{if } L_{G} < 0\\
    1  & \quad \text{else}
  \end{cases}
\end{equation}
More details about the reward function will be provided later in Section 3.4. 

The loss of each random walk will be multiplied by the corresponding reward to guide the network update such that the model tends to generate high-importance edges. We leverage this approach to generate edges that benefit the community detection task, even though the edge might not appear in the original network. The loss function of generator $G$ and discriminator $D$ are as follows:

\begin{equation}
L_{G} = {- \frac{1}{m}} \sum_{i=1}^{m}D(G(z^{(i)})) * f_{reward}(G(z^{(i)}))
\end{equation}

\begin{equation}
L_{D} = {\frac{1}{m}} \sum_{i=1}^{m}D(x^{(i)}) - \frac{1}{m} \sum_{i=1}^{m} D(G(z^{(i)}))
\end{equation}

In the next subsection, we will introduce how to train the GSGAN model.

\subsection{Training}
In Algorithm 1, we first sample a batch of noises from normal distribution as the generator's input. After the generator outputs random walks, we will calculate the score of each edge on this random walk, as shown in line 7 to 14. 
The goal here is to guide the generator to produce certain edges even when these edges receive low ratings from the discriminator (indicating that these edges do not exist in the original network), since these artificial edges are highly beneficial for community analysis tasks. We sum up the score of each edge as a reward for this random walk. In line 16, for the loss function, we multiply the loss obtained from each random walk by the corresponding reward. 
Even the discriminator gives a low rating to this random walk, the value may be raised after multiplying the reward. 
The advantage of this mechanism is that GSGAN can generate the edges which are important to the subsequent tasks (e.g., community detection), even though the edge may not exist in the original network. Next, we will introduce a measurement related to community detection and its improved version as our reward function.

\begin{algorithm}[htb] \label{algo. GSGAN}
\caption{GSGAN}  
\begin{algorithmic}[1] 
\State Given social network $SN = (V, E)$
\State Initialize $\theta_d$ for D and $\theta_g$ for G
\While {$\theta$ not converged}
    \State Sample $\{z^{(i)}|i=1,...,m\} \sim p(z)$ a batch of prior samples.
    \State Obtaining $Data_{gen}=\{G(z^{(i)})|i=1,...,m\}$
    \State $Reward = \{\}$ 
    \For{$G(z^{(i)})$ in $Data_{gen}$}
        \State $RW_{score} = 0$
        \For{$e$ in $G(z^{(i)})$}
            \State $Edge_{score} = f_{score(SN, e)}$
            \State $RW_{score}  = RW_{score} + Edge_{score}$
        \EndFor
        \State $Reward = concat(Reward, RW_{score})$
    \EndFor
    \State Update generator parameters $\theta_g$ to minimize 
        
        $\widetilde{V} = {- \frac{1}{m}} \sum_{i=1}^{m}D(G(z^{(i)})) * Reward^{(i)}$
        \State \qquad $\theta_g \gets \theta_g - \eta \nabla \widetilde{V}(\theta_g)$
    \State Sample $\{x^{(i)}|i=1,...,m\} \sim P_r$ a batch from the real data.
    \State Sample $\{z^{(i)}|i=1,...,m\} \sim p(z)$ a batch of prior samples.
    \State Update discriminator parameters $\theta_d$ to maximize 
    
    $\widetilde{V} = {\frac{1}{m}} \sum_{i=1}^{m}D(x^i) - \frac{1}{m} \sum_{i=1}^{m} D(G(z^{(i)}))$
        \State \qquad $\theta_d \gets \theta_d + \eta \nabla \widetilde{V}(\theta_d)$

\EndWhile  
\end{algorithmic}  
\end{algorithm}

\subsection{Reward}
Most methods for graph sparsification focus on retaining important edges among the existing ones, such as \cite{wong2017finding}\cite{satuluri2011local}\cite{hamann2016structure}. There are many ways to measure the importance of edges. For example, the Jaccard coefficient, edge betweenness centrality and clustering coefficient, but not all measurements are helpful for the community detection task. From the aspect of community detection, keeping intra edges in the community should have higher priority over inter edges between communities. 
A commonly accepted assumption is that if there are more common neighbors between two nodes, they are more likely to be connected by a link, and hence they also have a higher chance to be in the same community. 
Following this idea, we first introduce the Jaccard coefficient measurement and then propose a modified version for better evaluating the importance of an edge to the community detection task.

\underline{\textbf{Jaccard coefficient:}}
Jaccard coefficient is a measurement to compare the similarity and diversity of a sample set. 
Jaccard coefficient can be used to measure the similarity between two nodes as Equation \ref{eq:9}, where $N_u$ and $N_v$ denote the neighborhood of node $u$ and $v$, respectively. The intuition behind this measurement is that if these two nodes are highly similar, they are more likely to be in the same community, and hence the relationship between them should be relatively important. Therefore, no matter there is a connection between these two similar nodes in the original network or not, the relationship between them should be concretized in the sparsified network graph.

\begin{equation} \label{eq:9}
Jaccard(u, v) = \frac{|N_u\cap N_v|}{|N_u\cup N_v|}
\end{equation}

\underline{\textbf{Density Jaccard:}}
One problem of the aforementioned basic Jaccard coefficient is lacking the consideration of cluster density. In Fig. 2(c), there are two communities (i.e., the eight nodes on the top, and the four nodes in the bottom). When calculating the Jaccard coefficient of an edge $(u,v)$, the edge in the community above tends to be considered as more important than the edge in the community below. 
Therefore, using Jaccard coefficient to measure the importance of edges would easily cause the community below loosing most of its edges and becoming a bunch of isolated points. To address this problem, we improve the basic Jaccard coefficient to take the density into consideration. Let $G_N(u, v)$ be the subgraph formed by the neighbors of $u$ and $v$, its density is defined as $D(G_N(u, v)) = \frac{2|E_N(u, v)|}{|V_N(u, v)|(|V_N(u, v)|-1)}$, where $|V_N(u, v)|$ and $|E_N(u, v)|$ is the number of nodes and the number of edges in the subgraph $G_N(u, v)$, respectively. The Density Jaccard coefficient of an edge $(u,v)$ is then defined as follows:

\begin{equation}
Density Jaccard(u, v) = \frac{|N_u\cap N_v|}{|N_u\cup N_v|*D(G_N(u, v))}
\end{equation}

In GSGAN, the above measurements are used in the reward function to calculate the importance of edges. With these measurements, when the discriminator $D$ evaluates each edge in the random walk, it not only considers the correctness but also the relationship between the nodes at the two ends of that edge, and thus capturing the important relationships represented by the artificial edges. 

\subsection{Assembling Social Network}
Since the final output of GSGAN should be a sparsified network rather than some random walks, we still need to transform the generated random walks into a social network. 
First, the trained generator $G$ produces a large number of random walks. After counting the number of times each edge appearing in these random walks, we can obtain a matrix $S$, where each entry in $S$ represents the number of times that an edge appears. After dividing by the total number of edges, $S$ becomes a probability matrix. Next, GSGAN need to choose suitable edges based on $S$. A simple method is to set a threshold or directly select top-k edges, but this kind of approach may lead to the lost of low-degree nodes or even produce many singletons. Other studies also emphasize the importance of network connectivity \cite{zhou2010network}\cite{zhou2012simplification}. Therefore, we design a more sophisticated selection approach as described in Algorithm 2. 
Specifically, in line 1 to 3 of Algorithm 2, we first pick at least one edge for each node. Then, if the number of edges in the generated target graph $G_{T}$ is fewer than the desired number of edges $d$, we continue sampling the edge $(i,j)$ with probability $p_{ij}=\frac{S_{ij}}{\sum_{uv}S_{uv}}$, as described in line 4 to 6. Otherwise, if the number of edges already exceeds $d$, we will sequentially remove the edges according to the probability $p_{ij}$ (from low to high), as described in line 7 to 9.

\begin{algorithm} \label{algo. assembling_social_network}
\caption{Assembling Social Network}  
\begin{algorithmic}[1] 
\Require matrix $S$, desired number of edges $d$
\Ensure target network $G_{T}$
\For{$i = 1$ to $N$}
    \State $G_{T} = G_{T} \cup edge(i, \mathop{\arg\max}_{j}(\frac{S_{ij}}{\sum_{v}S_{iv}}))$
\EndFor
\While{number of edges in $G_{T} < d$}
    \State $G_{T} = G_{T} \cup edge(\mathop{\arg\max}_{ij}(\frac{S_{ij}}{\sum_{uv}S_{uv}}))$
\EndWhile
\While{number of edges in $G_{T} > d$}
    \State $G_{T} = G_{T}$ delete $edge(\mathop{\arg\min}_{ij}(\frac{S_{ij}}{\sum_{uv}S_{uv}}))$
\EndWhile
\State \Return $G_{T}$
\end{algorithmic}  
\end{algorithm}

\section{Experiments}
In this section, we evaluate and compare the proposed GSGAN with six baseline methods with three real-world datasets under three different clustering algorithms of community detection. Section 4.1 first describes the datasets, Section 4.2 then introduces the baselines and the evaluation metrics, and detailed analysis and discussions are provided in Section 4.3.

\subsection{Datasets}
We use three real-world datasets marked with ground-truth labels, and all of them are downloaded from the SNAP-dataset website. Details of these datasets are shown in Table I.
\begin{enumerate}
\item \textbf{Youtube social network} \cite{yang2015defining}: Youtube is a media sharing website where users can form friendships with each other, and users can also create groups for other users to join. We use the subgraph formed by the top-100 communities to evaluate the community detection performance.
\item \textbf{email-Eu-core network} \cite{leskovec2007graph}: This is an e-mail network, where an edge $(u,v)$ indicating that $u$ and $v$ have least one e-mail communication record. 
\item \textbf{Facebook} \cite{leskovec2012learning}: Facebook is a famous social media platform where users create various communities. This dataset is downloaded from the SNAP-dataset website, and different circles in the dataset are treated as different communities.
\end{enumerate}

\begin{table}[!htbp]
\caption{Dataset details. }
\centering
\begin{tabular}{cccc}
\hline
Dataset& Nodes& Edges& Communities\\
\hline
Youtube& 4890& 20787& 100\\
Eumail& 1005& 16064& 42\\
Facebook& 4039& 88234& 193\\
\hline
\end{tabular}
\end{table}

\subsection{Evaluation Metrics and Baselines}
The way we evaluate the performance of sparsification is to compare the difference between the community detection results obtained from the original network and the sparsified networks using the same clustering algorithm of community detection. If the results obtained from the sparsified network are very close (or even better) to the results from the original network, then this sparsification algorithm is considered as a good one. This sparsified network not only saves storage space but also accelerates the clustering algorithm of community detection, while maintaining the quality for community detection results. 
Because we use the datasets with ground-truth, we can adopt the external metric\footnotemark, Adjusted Rand Index (ARI) \cite{hubert1985comparing} as Equation \ref{eq:11}, to evaluate the quality of community detection results.

\footnotetext{There are two types of metrics for measuring community detection performance: external and internal. The data with ground-truth community labels use the external metric and data without ground-truth labels use the internal metric.}

\begin{equation} \label{eq:11}
ARI = \frac{\sum_{ij}\tbinom{n_{ij}}{2}-[ \,\sum_{i}\tbinom{a_i}{2}\sum_{j}\tbinom{b_j}{2}] \,/\tbinom{n}{2} }{\frac{1}{2}[ \,\sum_i\tbinom{a_i}{2}+\sum_j\tbinom{b_j}{2}] \,-[ \,\sum_i\tbinom{a_i}{2}\sum_j\tbinom{b_j}{2}] \,/\tbinom{n}{2}}
\end{equation}

\begin{figure*}[ht]
    \centering
    \includegraphics[width=1\textwidth]{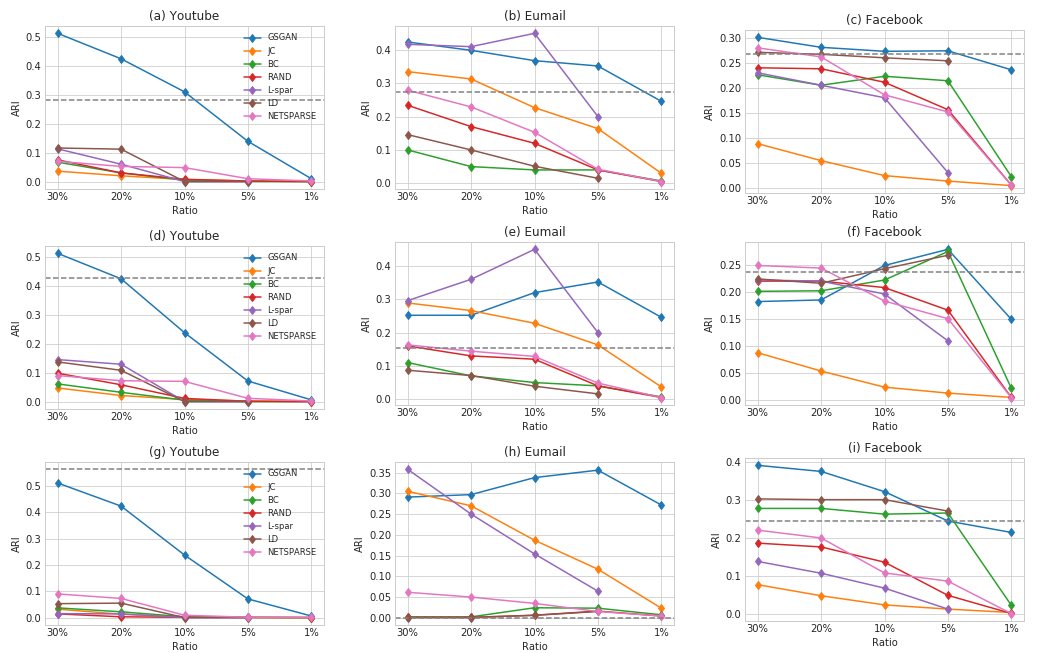}
    \caption{The performance of different sparsification methods under the different clustering algorithms. The first row represents the results of sparsification for Louvain algorithm. The second row represents the results of sparsification for Clauset algorithm. The third row represents the results of sparsification for the label propagation clustering algorithm.}
    \label{fig:fig3}
\end{figure*}

The compared baselines are listed as following:
\begin{enumerate}
\item Jaccard-coefficient-based approach (i.e., \textbf{JC}): In this method, we focus on the edges that exist in the network. We apply Jaccard coefficient similarity to measure the importance of each edge, and selecting top $r\%$ of edges to form the sparsified network $G'$. 
\item Edge-betweenness-centrality-based approach (i.e., \textbf{BC}): In addition to the Jaccard similarity, we further use edge betweenness centrality to measure the importance of edges, since it is also a commonly used measurement for the importance of edges. Betweenness centrality of an edge $e$ is the sum of the fraction of all-pairs shortest paths that pass through $e$. In Equation \ref{eq:12}, $\sigma(s, t)$ is the number of shortest $(s, t)$-paths, and $\sigma(s, t|e)$ is the number of those paths passing through the edge $e$. 
Those edges that are passed multiple times by the shortest paths are considered to be important bridges in the social network. We order the edges in the network according to the edge betweenness centrality and select top $r\%$ of the edges with highest values.
\begin{equation} \label{eq:12}
c_B(v) =\sum_{s,t \in V} \frac{\sigma(s, t|e)}{\sigma(s, t)}
\end{equation}
\item Random selection (i.e., \textbf{RAND}): There are many graph sparsification studies that also use random selection as a baseline, such as \cite{adityanetgist}. In this method, we randomly select $r\%$ of the edges in the network.
\item \textbf{L-spar} \cite{satuluri2011local}: The authors suggest that when doing sparsification, it is necessary to consider that different communities may have different densities. Only selecting the edges according to the Jaccard similarity from high to low may destroy the structure of the low-density community. 
Therefore, the authors propose L-Spar, selecting $degree^{r}$ edges for each node, so as to avoid the above problem.
\item \textbf{LD} \cite{hamann2016structure}: The goal of this approach is to keep those edges in the sparsified graph that lead to nodes with high degrees, i.e. the hubs that are crucial for a complex network's topology. The edges left after filtering can be considered a "hub backbone" of the network.
\item \textbf{NETSPARSE} \cite{gionis2017community}: This algorithm sparsifies the network to maintain the structural of a given community. The input of NETSPARSE consists of not only a network but also a set of communities. In addition, the process is performed using $\epsilon=0.5$, which is the harmonic mean of the average shortest path length and average degree in the sparsified graph.
\end{enumerate}

In the experiments, due to that L-spar, LD and NETSPARSE cannot accurately control the number of edges, we adjust the parameters of these methods to confirm that the numbers of edges in the sparsified graphs generated by these methods are consistent with the number of edges in our sparsified graph. Note that some methods cannot cover ratio 1 (i.e., the number of edges in the sparsified network is only 1\% of that number in the original network), so we only present the results up to their limits.

\begin{figure*}[ht]
    \centering
    \includegraphics[width=1\textwidth]{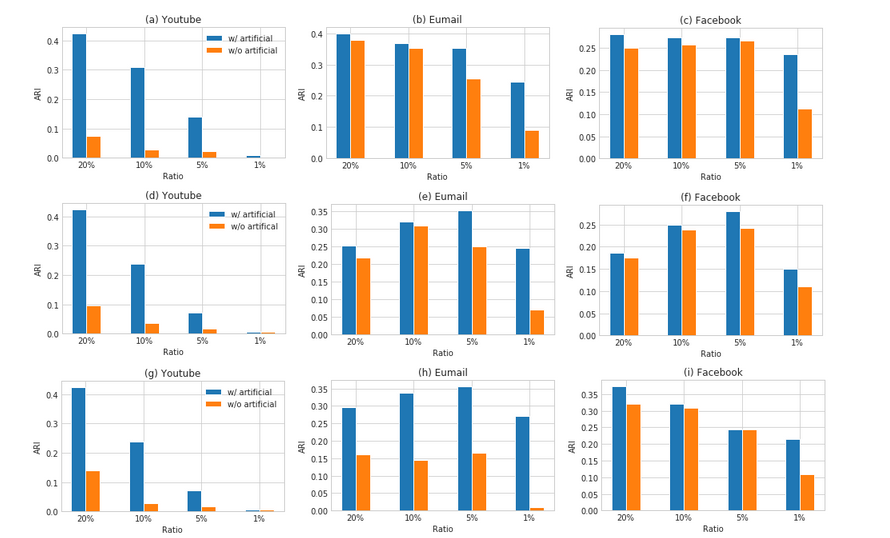}
    \caption{The performance of the graphs with and without artificial edges.}
    \label{fig:fig4}
\end{figure*}

\subsection{Results and Analysis}
In order to evaluate the effectiveness and flexibility of our proposed method, we validate our sparified graph $G'$ under three classic clustering algorithms of community detection, including Clauset\cite{clauset2004finding}, label propagation\cite{cordasco2010community}, and louvain\cite{blondel2008fast}. We also perform cross-comparisons with and without artificial edges and ablation tests, and we also show the time reduction of the community detection algorithms after sparsification. Finally, we visualize the sparsified graph.

\subsubsection{Comparison with baselines}
We present our experimental results in curve graphs, where different methods are distinguished by different colors. Note that the grey dotted line represents the \emph{standard line}, which is the accuracy of community detection using the original network without sparsification.

The first row of Fig. \ref{fig:fig3} indicates the performance of community detection in the Louvain algorithm. In Fig. \ref{fig:fig3}(a), it can be seen that our method defeats all the baselines. At ratio 20, our results even exceed the standard line, which means that the network after sparsification has better community detection performance than the original network. As we mentioned earlier, the real-world network tends to have noises, and the sparsification can help reduce these noises and bring better performance. At ratio 10, our method still maintains performance similar to the standard line, while other baselines have lost their usefulness.
The performance of our method in Fig. \ref{fig:fig3}(b) is higher than the standard line in all ratios. Although L-spar exceeds us at both ratio 20 and ratio 10, this may due to L-spar is based on the improvement of the Jaccard coefficient. It can be found that just a BC method can achieve good performance in this network, so L-spar also works well. However, when the ratio comes to 5 or even 1, our method gets a go-ahead. Note that even at ratio 1, our method still maintains the same performance as the original network, and other baselines almost cannot work at such a low ratio.
In Fig. \ref{fig:fig3}(c), our method on the Facebook network is still invincible. Although the BC, BC and RAND methods have certain effects at ratio 20 and ratio 10, they still below the standard line. It can be observed that until to ratio 1 our method is higher than the standard line. Note that, when the ratio comes to 1 our method still works, and other baselines have lost their effectiveness. In addition, although the effects of LD at ratio 20, ratio 10, and ratio 5 are higher than the standard line, they are still slightly lower than our method, and their method cannot be reduced to ratio 1, but our method still maintains good performance at ratio 1.

\begin{table*}[!ht]
\caption{The results of the ablation test for Louvain algorithm. We peel off the components layer by layer and compare the performance gap between them.}
\centering
\begin{tabular}{|c|c|c|c|c|c|c|c|c|c|c|c|c|}
\hline
\multirow{3}{*}{}      & \multicolumn{12}{c|}{Dataset}                                                                                                                                                                            \\ \cline{2-13} 
                       & \multicolumn{4}{c|}{Youtube}                                     & \multicolumn{4}{c|}{Eumail}                                       & \multicolumn{4}{c|}{Facebook}                                     \\ \cline{2-13} 
                       & 20\%           & 10\%           & 5\%            & 1\%           & 20\%           & 10\%           & 5\%            & 1\%            & 20\%           & 10\%           & 5\%            & 1\%            \\ \hline
GSGAN\_NRWs & 0.056          & 0.052          & 0.023          & 0.005         & 0.088          & 0.089          & 0.108          & 0.097          & 0.081          & 0.1            & 0.026          & 0.001          \\ \hline
GSGAN\_NR                 & 0.402          & 0.230          & 0.081          & 0.008         & 0.367          & 0.346          & 0.321          & 0.236          & 0.264          & 0.253          & 0.249          & 0.186          \\ \hline
GSGAN\_Jaccard          & 0.424          & 0.310          & 0.139          & 0.010         & 0.399          & 0.368          & 0.352          & 0.246          & 0.281          & 0.273          & 0.274          & \textbf{0.236} \\ \hline
GSGAN         & \textbf{0.431} & \textbf{0.314} & \textbf{0.184} & \textbf{0.11} & \textbf{0.409} & \textbf{0.386} & \textbf{0.375} & \textbf{0.266} & \textbf{0.293} & \textbf{0.283} & \textbf{0.275} & 0.225          \\ \hline
\end{tabular}
\end{table*}

\begin{figure*}[ht]
    \centering
    \includegraphics[width=1\textwidth]{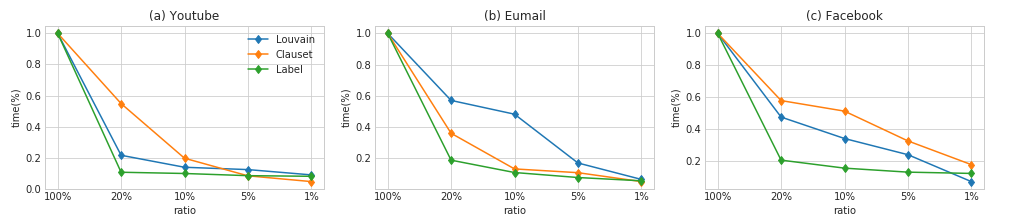}
    \caption{The execution time of clustering algorithm on sparsified networks.}
    \label{fig:fig6}
\end{figure*}

The second row indicates the performance of community detection in Clauset algorithm. In Fig. \ref{fig:fig3}(d), we can observe that only our method achieves the standard line at ratio 20, and the performance of other baselines on community detection is lower than ours at all ratios. Fig. \ref{fig:fig3}(e) also indicates that our method exceeds the standard line at all ratios. Although L-spar exceeds us at both ratio 20 and ratio 10, when comes to ratio 5 and ratio 1, our method surpasses them. An important fact is that only our method still works with ratio 1. In Fig. \ref{fig:fig3}(f), BC, LD and our method have similar performance at ratios 20, 10 and 5. But again, only our proposed method still works at ratio 1.

The third row indicates the performance of community detection in label propagation clustering algorithm. Although our method does not reach the standard line, it is still the most powerful in Fig. \ref{fig:fig3}(g). Moreover, Fig. \ref{fig:fig3}(h) shows that only BC and L-spar are close to us at ratio 20, but when the ratio is further reduced, all baselines are far below us. In Fig. \ref{fig:fig3}(i), BC and LD perform better than us only at ratio 5, and again, only our method is still working at ratio 1. According to the above experimental results, our sparsification method is flexible and effective, and it can be well applied to different clustering algorithms of community detection.

\subsubsection{Analysis of artificial edges}
In addition to comparing our methods with baselines, we also want to evaluate how helpful the artificial edges are to the community detection task. Hence, we conduct experiments for comparison of the sparsified networks with the artificial edges and without the artificial edges, as shown in Fig. \ref{fig:fig4}. Specifically, to generate a sparsified network without the artificial edges, we skip artificial edges and only choose edges that already exist in the network during the edge selection process.

Like the layout of Fig. \ref{fig:fig3}, the first row, second row, and third row in Fig. \ref{fig:fig4} show our experimental results in Louvain, Clauset, and label propagation clustering algorithms, respectively.
In Fig. \ref{fig:fig4}(a), it can be seen that after the artificial edges are removed, the performance collapses at all ratios. Fig. \ref{fig:fig4}(b) shows that the performance declines after the artificial edges are removed, and the performance drops as the ratio decreases. In Fig. \ref{fig:fig4}(c), performance without artificial edges is relatively low at all ratios, especially when the ratio is 1. Same as the experiments for the Louvain algorithm, Fig. \ref{fig:fig4}(d) shows that the performance is severely affected after removing artificial edges. Fig. \ref{fig:fig4}(e) shows that although it can still maintain a certain level of performance without using artificial edges, after adding artificial edges, the performance is improved, especially at ratio 5 and ratio 1. Fig. \ref{fig:fig4}(f) shows that the effectiveness has not improved significantly after adding the artificial edges. One possible reason is that the number of artificial edges is insufficient because those important relationships already have corresponding links in the original graph. The results of Fig. \ref{fig:fig4}(g)-\ref{fig:fig4}(i) have the same trend as previously discussed.



\subsubsection{Ablation tests}
In this subsection, we compare different versions of GSGAN as the ablation test, which are GSGAN without the real random walk, GSGAN without reward, GSGAN with Jaccard, and the proposed GSGAN.
\begin{enumerate}
\item \textbf{GSGAN without real random walk}: In order to verify the effectiveness of the random walk for graph sparsification. In GSGAN, we do not use the real random walk as training data to train the model, instead, we replace it with a path composed of randomly selected nodes. Moreover, we also applied a reward during training.
\item \textbf{GSGAN without reward}: In our proposed method GSGAN, we use the reward to train the model. In order to confirm the effectiveness of the reward, we remove the reward from GSGAN and observe how much performance will lose.
\item \textbf{GSGAN with Jaccard}: In an initial version of GSGAN, we measure the Jaccard similarity of the edges in the random walk as the reward. The trained model not only focuses on the existing edges and selects the important ones, but also captures those important relationships that do not exist in the original network. Therefore, this model is also able to create artificial edges useful for community detection. Note that this is just the initial version of GSGAN, we have improved the basic Jaccard similarity afterward.
\end{enumerate}
The experimental results are shown in Table II, and it can be seen that our proposed GSGAN (i.e., the final version with DensityJaccard reward) achieves the best results.

\subsubsection{Time reduction}
In addition to reducing the storage space, graph sparsification can also speed up the execution of the algorithm. Here, we evaluate the execution time of three classic clustering algorithms of community detection on our sparsified networks. 

As shown in Fig. \ref{fig:fig6}, the execution time of Louvain, Clauset and label propagation at ratio 20 of the Youtube social network is reduced by 78\%, 45\%, and 89\%, respectively. In the Eumail network, running Louvain, Clauset and label propagation at ratio 20 reduce the execution time by 42\%, 63\%, and 81\%, respectively. Finally, running Louvain, Clauset and label propagation at ratio 20 of Facebook social network reduce execution time by 52\%, 41\%, and 79\%, respectively.
\begin{figure*}[ht]
    \centering
    \includegraphics[width=1\textwidth]{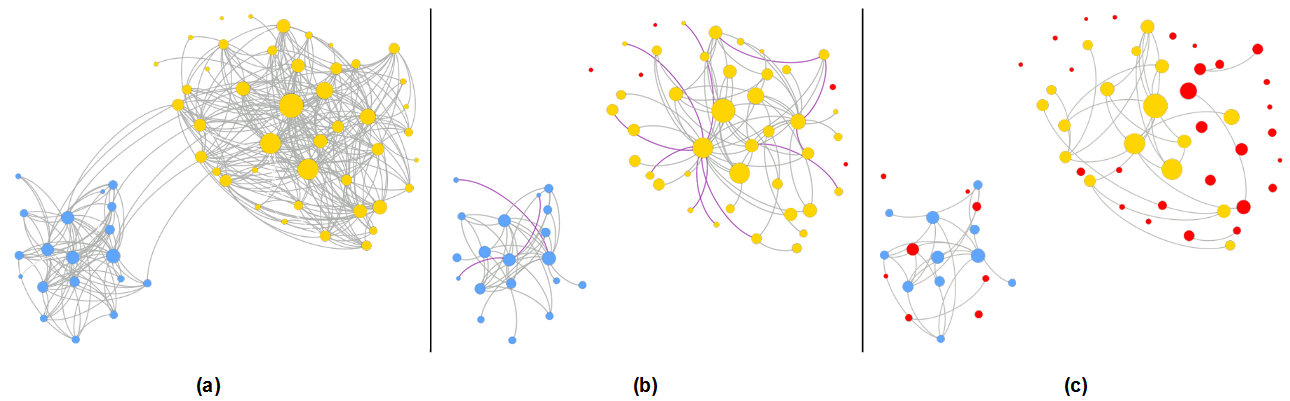}
    \caption{(a) represents the original graph with ground-truth community label. (b) and (c) represent the community detection results on sparsified graphs obtained by GSGAN and JC, respectively. Blue and yellow nodes indicate different communities. Purple edges represent the artificial edges generated by GSGAN, and the red nodes represent the nodes that are not in the correct community.}
    \label{fig:fig8}
\end{figure*}
\subsubsection{Graph visualization}
In the Fig. \ref{fig:fig8}, it can be found that GSGAN tends to eliminate the edges between the communities, and the artificial edges are used to make the structure of the community more compact. Although JC still roughly separates the two communities, the community detection using sparsified networks generated by JC does not perform well. The red nodes in the figure represent the nodes that are not in the correct community. On the contrary, the community detection with the sparsified graphs generated by GSGAN is able to locate most of the nodes in the correct community. It should be noted that the number of edges in Fig. \ref{fig:fig8}(b) seems to be larger than Fig. \ref{fig:fig8}(c), but in the entire sparsified graph, the total number of edges selected by GSGAN and JC are the same. That is, GSGAN is able to spend most of the quota on the edges inside the community (rather than the edges between the communities).

\section{Conclusions}
In this study, we propose a novel graph sparsification algorithms called GSGAN. Our proposed method uses random walks to capture the structure of networks and retain edges that are important for the community detection task. In addition to preserving essential edges from the original network, our method also considers the edges that do not exist in the original network but still contain important information. That is, GSGAN is able to create and add artificial edges to the sparsified network in order to further improve the performance of community detection. We evaluate the proposed method and other baselines using three real-world social networks under three clustering algorithms of community detection. The experimental results show that our method is quite effective and efficient under various community detection algorithms, especially at extremely low ratios.

\bibliographystyle{IEEEtran}
\bibliography{mybib}

\end{document}